\documentclass[8.5pt,twoside,twocolumn]{article}
\usepackage{mhchem}
\usepackage{times,mathptmx}
\usepackage{sectsty}
\usepackage{balance} 
\usepackage{graphicx}
\usepackage{lastpage}
\usepackage[format=plain,justification=raggedright,singlelinecheck=false,font=small,labelfont=bf,labelsep=space]{caption} 
\usepackage{amsmath}
\usepackage{amssymb}
\usepackage{upgreek}
\usepackage{dblfloatfix}

\DeclareMathAlphabet{\mathcal}{OMS}{cmsy}{m}{n}

\usepackage{xparse}
\ExplSyntaxOn
\NewDocumentCommand{\mref}{m}{\quinn_mref:n {#1}}
\seq_new:N \l_quinn_mref_seq
\cs_new:Npn \quinn_mref:n #1
 {\seq_set_split:Nnn \l_quinn_mref_seq { , } { #1 }
  \seq_pop_right:NN \l_quinn_mref_seq \l_tmpa_tl
  (\seq_map_inline:Nn \l_quinn_mref_seq
    { \ref{##1},\nobreakspace } 
  \exp_args:NV \ref \l_tmpa_tl)
 }
\ExplSyntaxOff
\usepackage{mathtools}
\usepackage{textcomp}
\usepackage{fancyhdr}
\begin{document}
\thispagestyle{plain}
\fancypagestyle{plain}{
\renewcommand{\headrulewidth}{1pt}}
\renewcommand{\thefootnote}{\fnsymbol{footnote}}
\renewcommand\footnoterule{\vspace*{1pt}% 
\hrule width 3.4in height 0.4pt \vspace*{5pt}} 
\setcounter{secnumdepth}{5}
\makeatletter 
\def\subsubsection{\@startsection{subsubsection}{3}{10pt}{-1.25ex plus -1ex minus -.1ex}{0ex plus 0ex}{\normalsize\bf}} 
\def\paragraph{\@startsection{paragraph}{4}{10pt}{-1.25ex plus -1ex minus -.1ex}{0ex plus 0ex}{\normalsize\textit}} 
\renewcommand\@biblabel[1]{#1}            
\renewcommand\@makefntext[1]% 
{\noindent\makebox[0pt][r]{\@thefnmark\,}#1}
\makeatother 
\renewcommand{\figurename}{\small{Fig.}~}
\sectionfont{\large}
\subsectionfont{\normalsize} 

\renewcommand{\headrulewidth}{1pt} 
\renewcommand{\footrulewidth}{1pt}
\setlength{\arrayrulewidth}{1pt}
\setlength{\columnsep}{6.5mm}

\twocolumn[
  \begin{@twocolumnfalse}
\noindent\LARGE{\textbf{From adhesion to wetting of a soft particle}}
\vspace{0.6cm}

\noindent\large{\textbf{Thomas Salez, Michael Benzaquen, and Elie Rapha\"el}}\vspace{0.5cm}\\
\noindent\textit{\small{\textbf{Received Xth XXXXXXXXXX 20XX, Accepted Xth XXXXXXXXX 20XX\newline
First published on the web Xth XXXXXXXXXX 20XX}}}
\noindent \textbf{\small{DOI: 10.1039/b000000x}}
\vspace{0.6cm}

\noindent \normalsize{\textit{Using a thermodynamical approach, we calculate the deformation of a spherical elastic particle placed on a rigid substrate, under zero external load, and including an ingredient of importance in soft matter: the interfacial tension of the cap.  In a first part, we limit the study to small deformation. In contrast with previous works, we obtain an expression for the energy that precisely contains the JKR and Young-Dupr\'e asymptotic regimes, and which establishes a continuous bridge between them. In a second part, we consider the large deformation case, which is relevant for future comparison with numerical simulations and experiments on very soft materials. Using a fruitful analogy with fracture mechanics, we derive the exact energy of the problem and thus obtain the equilibrium state for any given choice of physical parameters.}}
\vspace{0.5cm}
 \end{@twocolumnfalse}
  ]
\footnotetext{\textit{Laboratoire de Physico-Chimie Th\'eorique, UMR CNRS Gulliver 7083, ESPCI, Paris, France. }}
Since the seminal works of Hertz~\cite{Hertz1896,Timoshenko1970}, Johnson, Kendall and Roberts (JKR)~\cite{Johnson1971}, and Derjaguin, Muller and Toporov (DMT)~\cite{Derjaguin1975}, the contact of adhesive elastic solids has been widely studied~\cite{Johnson1985,Maugis2000,Barthel2008}. This area of research is of tremendous importance: the range of application now spreads from biology to engineering, as shown by the recent developments on latex particles~\cite{Lau2002}, biological cells~\cite{Brochard2003,Chu2005}, or micro-patterned substrates~\cite{Poulard2011} for instance. Extensions of JKR theory to large deformation have been obtained, and the JKR-DMT transition has been clarified: see for instance Maugis' Book~\cite{Maugis2000} for an interesting historical review on the topic. In the Boussinesq problem of an infinite elastic half-space indented by a rigid sphere, exact theories have been proposed~\cite{Maugis1995} using Sneddon theorems~\cite{Sneddon1965}. The dual problem of an elastic sphere on a rigid substrate has been studied as well in symmetric compression~\cite{Tatara1991}. In a similar way, wetting properties have been subject to an abundant literature~\cite{deGennes1985,deGennes2003,Bonn2009}. Wetting on elastic substrates has been intensely studied~\cite{Py2007,Duprat2011,Marchand2012,Style2012} and electrowetting~\cite{Pollack2000,Mugele2005} now allows to precisely control the wetting properties of a soft material.

\begin{figure}[t!]
\begin{center}
\includegraphics[width=7cm]{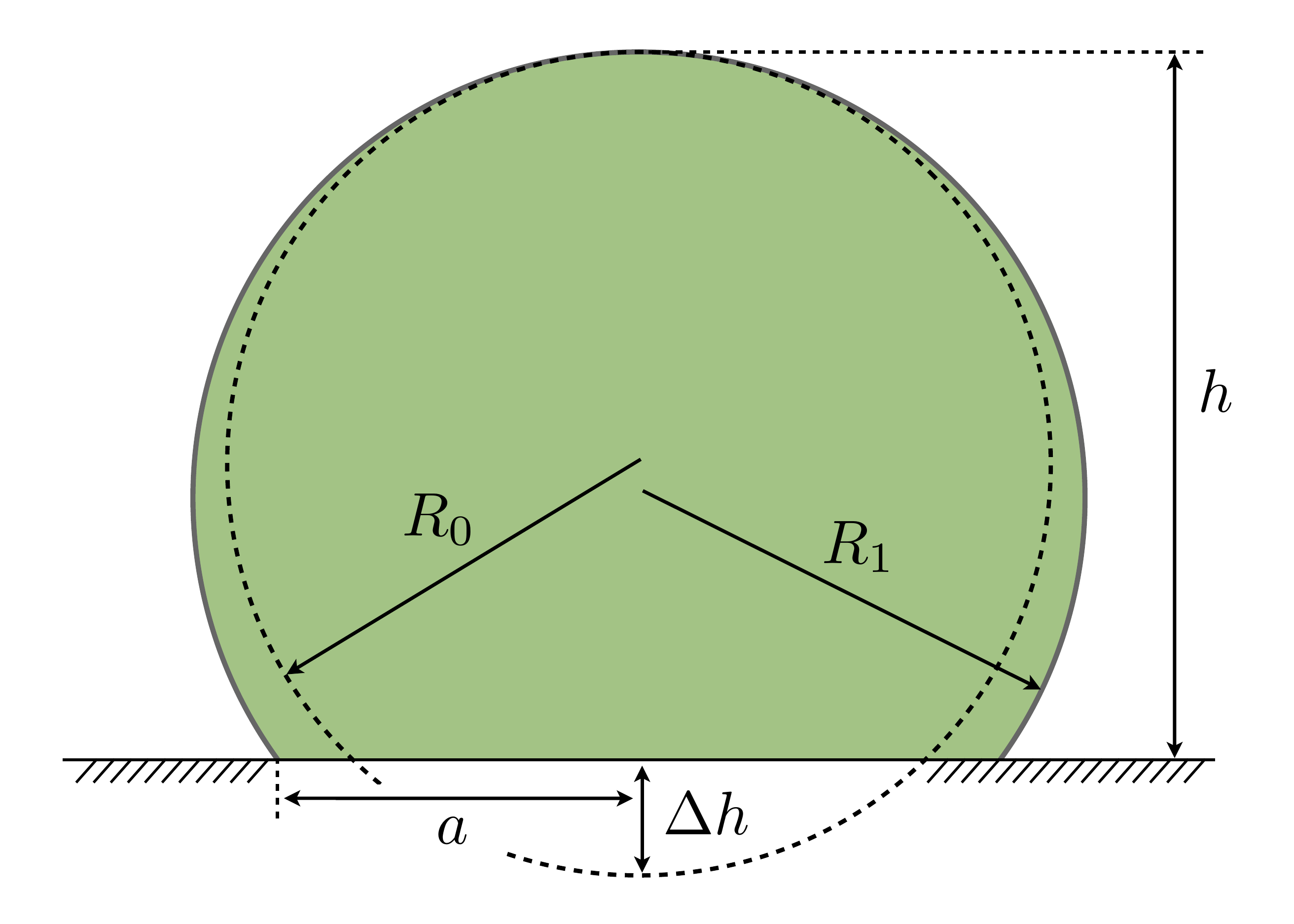}
\caption{\textit{Schematic of a soft spherical particle of initial radius $R_0$ (dashed line) deposited onto a rigid substrate. After capillary spreading, the deformed state is characterized by the radius $R_1$ of the external spherical cap. Note that this simple picture does not account for the actual deformation of the particle at the edges of the contact zone\cite{Maugis2000}.}}
\label{fig1}
\end{center}
\end{figure}

In the present article, we thermodynamically calculate the shape of a spherical elastic particle on a single rigid substrate (see Fig.~\ref{fig1}), under zero external load, and we include an ingredient of importance in soft matter~\cite{Lau2002,Carrillo2010,Carrillo2012}: the interfacial tension of the cap, that was neglected in JKR theory although capillary adhesion was taken into account. This supplementary ingredient allows one to draw a bridge between adhesion and wetting for any soft object. 
In a first part, we limit the study to small deformation. In previous static~\cite{Lau2002,Carrillo2010} and dynamic~\cite{Carrillo2012} works, the choice was made on obtaining proper scalings rather than the two exact asymptotic behaviours, namely the Young-Dupr\'e and JKR theories. This choice was valid in view of the spherical shape assumption, which is only approximate near the edges of contact~\cite{Maugis2000}. However, these attempts remain at the level of scaling and thus do not allow for quantitative description of experiments and numerical simulations. Here, in contrast, we build up an expression for the energy at small deformation which precisely contains the JKR and Young-Dupr\'e asymptotic regimes, thus establishing a continuous crossover between them. We stress that the main hypothesis underlying this new analysis is that the spherical shape is a good approximation when it comes to calculate the capillary energetic contribution of the external cap, but not for the elastic contribution itself for which we keep the exact JKR expression. In a second part, we consider the large deformation case. Using a fruitful analogy with fracture mechanics, as developed by Maugis for the dual Boussinesq problem~\cite{Maugis1995}, we obtain for the first time the exact energy and the equilibrium shape for any given choice of  physical parameters.

We consider a soft spherical elastic particle of initial radius $R_0$ that is deposited onto a rigid substrate (see Fig.~\ref{fig1}). Adhesion forces tend to increase the particle-substrate contact area, while the particle-vapour surface tension and bulk elasticity limit this process. In order to estimate the contribution of the particle-vapour surface tension, we assume that the equilibrium shape can be described as a spherical cap with radius $R_1$.
Incompressibility imposes:
\begin{equation}
\label{radius}
R_1=\frac{4R_0^{\,3}}{3h^2}+\frac{h}{3}\ .
\end{equation} 
In addition, due to spherical geometry, the contact radius $a$ is given by:
\begin{equation}
\label{contact}
a=\sqrt{2R_1h-h^2}\ .
\end{equation} 
Finally, we introduce the deformation depth $\Delta h$:
\begin{equation}
\label{defo}
\Delta h=2R_0-h\ ,
\end{equation}
as a unique variable.

\section{Model at small deformation}
In this first part, we present the Tenso-Elastic-Adhesive (TEA) model at small deformation: $\Delta h\ll  R_0$. Then, we obtain the analytical solution and compare it to the JKR and Young-Dupr\'e asymptotic regimes. Finally, we retrieve the results thanks to a fruitful analogy with fracture mechanics~\cite{Maugis1995} and compare them to previous expressions~\cite{Lau2002,Carrillo2010}.

Going beyond JKR theory, we wish to include the contribution of the surface tension of the external spherical cap. Therefore, we calculate the total TEA energy under zero external load:
\begin{equation}
\label{total}
U_{\textrm{TEA}}=U_{\textrm{ad}}+U_{\textrm{el}}+U_{\textrm{s}}\ ,
\end{equation} 
where $U_{\textrm{ad}}$, $U_{\textrm{s}}$, and $U_{\textrm{el}}$ are the adhesive, tensile and elastic energetic contributions at small deformation, respectively. According to JKR theory~\cite{Johnson1971}, the elastic energy at small deformation, and under zero external load, equals:
\begin{equation}
\label{elas}
U_{\textrm{el}}=\frac{1}{15}KR_0^{\,-2}a^5\ ,
\end{equation} 
with the rigidity\footnote{In previous works~\cite{Lau2002,Carrillo2010}, the rigidity was defined as $K=\frac{E}{1-\nu^2}=\frac{2G}{1-\nu}$.}: 
\begin{equation}
\label{rigi}
K=\frac{4}{3}\frac{E}{1-\nu^2}\ ,
\end{equation} 
where $E$ is the Young Modulus, and $\nu$ the Poisson ratio. Equation~\mref{elas} corresponds to a restoring energy that causes a resistance to the deformation. Using Eqs.~\mref{radius,contact,defo}, and developing Eq.~\mref{elas} at the lowest order in $\Delta h/R_0$, leads to\footnote{There is a $16/21$ factor in comparison with the developed elastic energy in previous works~\cite{Lau2002,Carrillo2010}. It is due to the use of the spherical connexion $\Delta h(a)$, obtained from Eqs.~\mref{radius,contact,defo} in the elastic energy from the dual Boussinesq problem, instead of the real Boussinesq connexion~\cite{Maugis2000}, and to a different but self-consistent definition of the rigidity $K$ (see previous foonote).}:
\begin{equation}
\label{elasfin}
U_{\textrm{el}}\approx\frac{4\sqrt{2}}{15}KR_0^{\,1/2}\Delta h^{5/2}\ .
\end{equation} 
The adhesive energy is given by:
\begin{equation}
\label{workad}
U_{\textrm{ad}}=-\pi Wa^2\ ,
\end{equation} 
where $W$ is the thermodynamical work of adhesion between the spherical particle (P) and the solid substrate (S), in the ambient vapor (V):
\begin{equation}
\label{W}
W=\gamma+\gamma_{\,\textrm{SV}}-\gamma_{\,\textrm{PS}}\ ,
\end{equation} 
with the notation $\gamma=\gamma_{\,\textrm{PV}}$. Note that it is straightforward in the spherical case to recover Eq.~\mref{workad} -- that depends only on the area of contact -- by integrating all the volumic van der Waals interactions between the two considered bodies~\cite{Israelachvili1992}. Equation~\mref{workad} expresses the fact that a positive adhesive work tends to deform the particle, by spreading. Using Eqs.~\mref{radius,contact,defo,workad} and removing the additional constant term gives:
\begin{equation}
\label{adfin}
U_{\textrm{ad}}\approx -2\pi WR_0\Delta h\ ,
\end{equation} 
at the lowest order in $\Delta h/R_0$. As in the case of elasticity (see Eq.~\mref{elas}), surface tension acts a restoring energy:
\begin{subequations}
\begin{align}
\label{tense1}
U_{\textrm{s}}&=\pi\gamma(a^2+2R_1h)\\
&=2\pi\gamma\left(a^2+\frac{h^2}{2}\right)\ ,
\end{align}
\end{subequations} 
according to Eq.~\mref{contact}, where we recognize the total surface of a spherical cap (see Fig.~\ref{fig1}). Note that we do not count twice the particle-substrate interaction since, according to Eqs.~\mref{workad,W,tense1} we have:
\begin{equation}
U_{\textrm{ad}}+U_{\textrm{s}}=\pi a^2 (\gamma_{\,\textrm{PS}}-\gamma_{\,\textrm{SV}})+2\pi \gamma R_1h\ ,
\end{equation} 
where the first term is the energetic cost of replacing the solid-vapor interface by the particle-substrate interface, through capillary adhesion, and the second term is the surface energy of the external spherical cap. Using Eqs.~\mref{radius,contact,defo}, and developing Eq.~(\ref{tense1}) at the lowest order in $\Delta h/R_0$, leads to:
\begin{equation}
\label{tense}
U_{\textrm{s}}\approx\pi\gamma\Delta h^{2}\ .
\end{equation} 
As introduced in Eq.~\mref{total}, the TEA energy is the sum of Eqs.~\mref{elasfin,adfin,tense}:\small
\begin{equation}
\label{ftot}
U_{\textrm{TEA}}\approx-2\pi WR_0\Delta h+\pi\gamma\Delta h^{2}+\frac{4\sqrt{2}}{15}KR_0^{\,1/2}\Delta h^{5/2}\ ,
\end{equation}\normalsize
at the lowest order in $\Delta h/R_0$. Then, let us introduce the dimensionless quantities:
\begin{subequations}
\begin{align}
\label{X}
X&=\frac{\gamma}{W}\\
\label{Y}
Y&=\frac{\Delta h}{2R_0}\\
\label{Z}
Z&=\frac{KR_0}{4W}\\
\tilde{U}_{\textrm{TEA}}&=\frac{U_{\textrm{TEA}}}{2\pi WR_0^{\,2}}\ .
\label{utildetea}
\end{align}
\end{subequations}
Finally, dividing Eq.~\mref{ftot} by $2\pi WR_0^{\,2}$, one gets the dimensionless expression of the total energy:
\begin{equation}
\label{ftotbis}
\tilde{U}_{\textrm{TEA}}\approx-2Y+2XY^2+\frac{64}{15\pi}ZY^{5/2}\ ,
\end{equation} 
at the lowest order in $Y$ that contains all the ingredients of the model (through $X$ and $Z$).

The Young-Dupr\'e regime corresponds to a non-elastic particle. Therefore, in order to study this limit, we set $Z=0$ in Eq.~\mref{ftotbis} according to Eq.~\mref{Z}. At constant temperature and volume, the thermodynamical equilibrium is reached when $\tilde{U}_{\textrm{TEA}}$ is minimal with respect to $Y$. Then, minimizing Eq.~\mref{ftotbis} with respect to $Y$ leads to the solution:
\begin{equation}
\label{ydsmall}
Y^*_{\textrm{YD}}=\frac{1}{2X}\ .
\end{equation} 
Note that the necessary condition $Y^*_{\textrm{YD}}<1$ is ensured, since one has $X>1/2$ in partial wetting.
Using Eqs.~\mref{W,X,Y}, one then obtains:
\begin{equation}
\frac{\Delta h^*_{\textrm{YD}}}{R_0}=1+\frac{\gamma_{\,\textrm{SV}}-\gamma_{\,\textrm{PS}}}{\gamma}\ .
\end{equation} 
The cosine of the equilibrium contact angle $\theta^*$ at small deformation is thus given by:
\begin{subequations}
\begin{align}
\cos\theta^*&\approx\frac{\Delta h^*_{\textrm{YD}}}{R_0}-1\\
&=\frac{\gamma_{\,\textrm{SV}}-\gamma_{\,\textrm{PS}}}{\gamma}\ ,
\label{it}
\end{align}
\end{subequations} 
which corresponds to Young-Dupr\'e law~\cite{deGennes2003}.

As already mentioned in the introduction of this article, JKR theory neglects the interfacial tension $\gamma$ of the particle in the considered atmosphere. Therefore, in order to study this limit, we set $X=0$ in Eq.~\mref{ftotbis} according to Eq.~\mref{X}. Minimizing Eq.~\mref{ftotbis} with respect to $Y$ leads to the solution:
\begin{equation}
\label{jkrsmall}
Y^*_{\textrm{JKR}}=\left(\frac{3\pi}{16Z}\right)^{2/3}\ .
\end{equation} 
Note that the necessary condition $Y^*_{\textrm{JKR}}<1$ is ensured as soon as $Z>3\pi/16$.
Finally, using Eqs.~\mref{radius,contact,defo,Y,Z,jkrsmall} one finds the JKR contact radius:
\begin{equation}
\label{jkrad}
a_{\textrm{JKR}}=\left(\frac{6\pi WR_0^{\,2}}{K}\right)^{1/3}\ ,
\end{equation} 
which is precisely\footnote{Removing the surface tension term in the developed total energy in previous work~\cite{Carrillo2010}, and looking for equilibrium, does not give back the exact JKR radius of Eq.~\mref{jkrad} due to the $16/21$ factor in the elastic term (see previous footnote). This is fully understood since JKR theory is incompatible with an imposed spherical shape, thus this previous study~\cite{Carrillo2010} was limited to scaling only by construction.} the JKR contact radius under zero external load~\cite{Johnson1971}.

\begin{figure}
\begin{center}
\includegraphics[width=8.7cm]{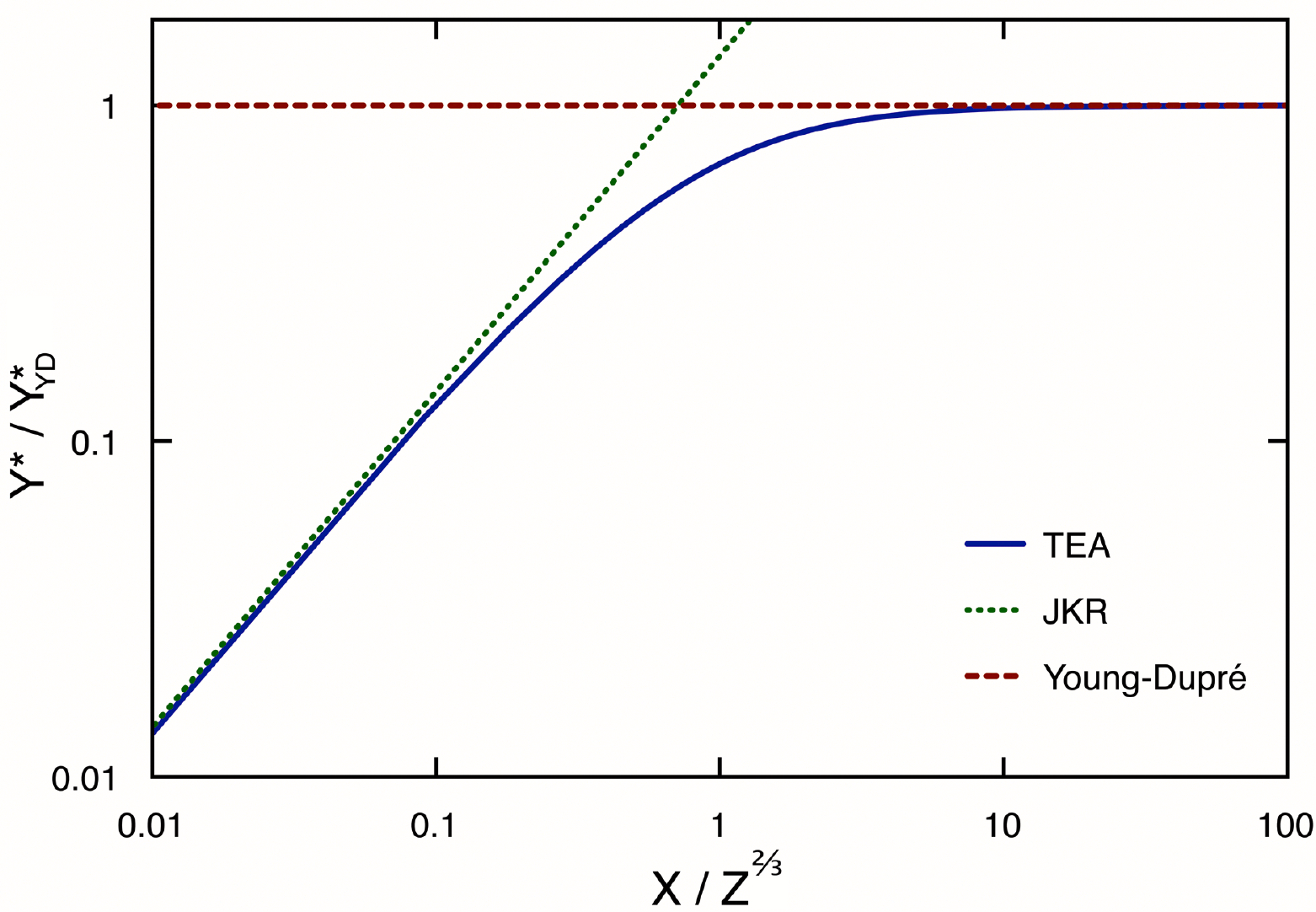}
\caption{\textit{Normalized solution of the TEA model to the Young-Dupr\'e one, from Eq.~\mref{solin}. For comparison, we plotted the Young-Dupr\'e ($Z=0$) and JKR ($X=0$) regimes at small deformation, according to Eqs.~\mref{ydsmall,jkrsmall}. The definitions of the dimensionless variables $X$, $Y$ and $Z$ are given in Eqs.~\mref{X,Y,Z}.}}
\label{fig2}
\end{center}
\end{figure}

Let us now consider the general case with finite $X$ and $Z$. The thermodynamical equilibrium is reached when $\tilde{U}_{\textrm{TEA}}$ is minimal with respect to $Y$, which corresponds to a cubic equation. Its positive real solution $Y^*$ is given by:
\begin{subequations}
\begin{align}
\frac{Y^*}{Y^*_{\textrm{YD}}}=\beta\left[\sqrt[3]{r+\sqrt{r^2+q^3}}+\sqrt[3]{r-\sqrt{r^2+q^3}}-\frac{\beta}{3}\right]^2\ ,
\label{solin}
\end{align}
\end{subequations}
\normalsize
where for clarity we have introduced the auxiliary variables:
\begin{subequations}
\begin{align}
\beta&=\frac{Y^*_{\textrm{JKR}}}{Y^*_{\textrm{YD}}}=\left(\frac{9\pi^2}{32}\right)^{1/3}\frac{X}{Z^{2/3}}\\
q&=-\left(\frac{\beta}{3}\right)^2\\
r&=\frac{1}{2}-\left(\frac{\beta}{3}\right)^3\ .
\end{align}
\end{subequations}
From Eq.~\mref{solin}, one can plot the normalised solution as a function of the \textit{unique} variable $\beta\propto XZ^{-2/3}$, as shown in Fig.~\ref{fig2}.
The TEA solution always leads to a smaller deformation than the single JKR or Young-Dupr\'e one due to the additional restoring energy. Moreover, we observe a smooth transition between the JKR and Young-Dupr\'e asymptotic regimes, the governing model being the one leading to the smaller deformation. This fundamental crossover should be experimentally observable when $Y^*_{\textrm{YD}}\approx Y^*_{\textrm{JKR}}$ that is for $\gamma^{\,3}\approx WK^{2}R_0^{\,2}$, within a one order of magnitude typical range, as long as the deformation remains small for the previous calculation to be valid. Note that if  $\Delta h\ll R_0$ is no longer satisfied, one will need a large deformation model.

To conclude this part on small deformation, we recall the analogy with fracture mechanics that was originally developed by Maugis for the dual Boussinesq problem~\cite{Maugis1995}. Then, we show that it gives back the correct elastic energy for our system at small deformation, and we explain the difference with the elastic energy obtained in previous studies~\cite{Lau2002,Carrillo2010}. In the Boussinesq canonical example, a hookean elastic infinite half-space is indented by a rigid sphere in presence of capillary adhesion and without surface tension of the external cap. In the field of fracture mechanics, it is well known that the elastic energy is fully released during the fracture process. Therefore, by analogy, the equilibrium between adhesion and elasticity is reached when the work of adhesion $W$ balances the fracture energy release rate $\mathcal{G}$:
\begin{subequations}
\begin{align}
W&=\mathcal{G}\\
&=\frac{a^3K}{6\pi R_0^{\,2}}\ ,
\label{frac}
\end{align}
\end {subequations}
where we wrote the expression of $\mathcal{G}$ at small deformation and under zero external load~\cite{Maugis1995}. As one can immediately see, this gives back the JKR contact radius of Eq.~\mref{jkrad}, which validates the analogy with fracture mechanics. Furthermore, Eq.~\mref{frac} is similar to a force balance. Let us integrate the two sides over the contact area, in order to obtain the elastico-adhesive energy. The adhesive term in Eq.~\mref{frac} gives, with appropriate sign:
\begin{equation}
U_{\textrm{ad}}=-\int_{\Sigma_{\textrm{contact}}}d\Sigma\  W\ ,
\end {equation}
where we retrieve precisely Eq.~\mref{workad}. The elastic term in Eq.~\mref{frac} gives, with appropriate sign:
\begin{equation}
\label{elasfinbis}
U_{\textrm{el}}=\int_{\Sigma_{\textrm{contact}}}d\Sigma\ \mathcal{G}\ ,
\end {equation}
where we retrieve precisely Eq.~\mref{elas}, and thus Eq.~\mref{elasfin} at small deformation.

For comparison, the elastic energy can also be evaluated using the fundamental hookean energy~\cite{Landau1990}:
\begin{subequations}
\begin{align}
U_{\textrm{el}}&=\frac{1}{2}\int_{V}dV\ \sigma_{ij} \epsilon_{ij}\\
&=\frac{1}{2}\oint_{\Sigma}d\Sigma\ \sigma_{ij} u_in_j\ ,
\label{second}
\end{align}
\end {subequations}
where $V$ and $\Sigma$ are the volume and surface before deformation, $\sigma_{ij}$ and $\epsilon_{ij}=(\partial_i u_{j}+\partial_j u_{i})/2$ are the components of the stress and strain symmetric tensors,  and $u_i$ is the local deformation along $i$. To obtain Eq.~\mref{second}, we used the internal equilibrium $\partial_j\sigma_{ij}=0$, and the Green-Ostrogradski theorem. In previous works~\cite{Lau2002,Carrillo2010}, only the vertical stress and strain from the Boussinesq problem~\cite{Maugis1995} are considered at the contact, thus:
\begin{equation}
\label{landausur}
U_{\textrm{el}}=\frac{1}{2}\int_{\Sigma_{\textrm{contact}}}d\Sigma\ \sigma_{zz} u_z\ ,
\end{equation} 
where the integral is evaluated over the coordinates of the system before deformation. Equation~\mref{landausur} should be identical to Eq.~\mref{elasfinbis}, since $\mathcal{G}$, $u_z$ and $\sigma_{zz}$, all come from the same analysis~\cite{Maugis1995}. However, in the Boussinesq problem, the correct total deformation $\delta(a)$ satisfies~\cite{Maugis2000}:
\begin{equation}
\label{delta}
\delta=\frac{a^2}{3R_0}\ ,
\end{equation} 
which means that $\delta \neq \Delta h$. Actually, using Eqs.~\mref{radius,contact,defo} at small deformation, one obtains:
\begin{equation}
\label{spherecon}
\Delta h\approx\frac{a^2}{2R_0}\ .
\end{equation} 
In previous works~\cite{Lau2002,Carrillo2010}, $\delta$ was directly replaced by the spherical connexion $\Delta h(a)$ from Eq.~\mref{spherecon} in the expressions of $u_z$ and $\sigma_{zz}$, thus leading to a wrong JKR contact radius. When considering instead Eq.~\mref{delta} inside the expressions of $u_z$ and $\sigma_{zz}$ given by Maugis~\cite{Maugis1995}, we get from Eq.~\mref{landausur}: 
\begin{equation}
\label{phi}
U_{\textrm{el}}=\frac{3}{4}K\left(a\delta^2-\frac{2}{3}\frac{a^3\delta}{R_0}+\frac{1}{5}\frac{a^5}{R_0^{\,2}}\right)\ ,
\end{equation}
which is equal to Eq.~\mref{elas}, and thus to Eq.~\mref{elasfin} at small deformation. This expression gives back the correct JKR radius of Eq.~\mref{jkrad} when minimizing the elastico-adhesive energy. An equivalent way to understand this difference is to notice that there are two ways of making the analogy with the Boussinesq problem. On one hand, it has been considered~\cite{Lau2002,Carrillo2010} that $\delta=\Delta h$ and $a$ are dependent variables from the beginning, \textit{i.e.} in Eq.~\mref{phi}, through the spherical connexion of Eq.~\mref{spherecon}. Therefore, the exact JKR result of Eq.~\mref{jkrad} cannot be obtained but the scaling is correct. On the other hand, the present TEA model starts from two independent variables, $\delta$ and $a$, in the Boussinesq energy of Eq.~\mref{phi}. The connexion of Eq.~\mref{delta} is then obtained by minimizing Eq.~\mref{phi} with respect to $\delta$ at constant $a$, and introduced back in Eq.~\mref{phi} thus leading to Eq.~\mref{elas}. Therefore, the TEA model starts with an elastic energy that depends only on $a$. This \textit{ad-hoc} approximation has the great advantage of containing the exact JKR contact radius of Eq.~\mref{jkrad}, and thus to allow for a quantitative comparison with experiments, even though we approximate the shape by a purely spherical cap. The main argument in favour of the new approach presented here is that a spherical cap gives the good estimate of the tensile energy of the external cap, and thus allow for Young-Dupr\'e limit to be reached, and at the same time the JKR elasticity gives the proper elastic contribution, and thus allow for the JKR limit to be reached (see Fig.~\ref{fig2}).
\newpage

\section{Model at large deformation}
To understand experiments or numerical simulations that reach large deformation, one cannot use the small deformation energy of Eq.~\mref{ftotbis}. Therefore, one needs a theory at large deformation. In this second part, we thus extend the previous analogy which fracture mechanics to large deformation by using the exact energy release rate obtained by Maugis for the dual Boussinesq problem~\cite{Maugis1995}. Note that \textit{large deformation} means that we do not restrict ourselves anymore to an approximate parabolic shape around the contact zone, but we use the exact spherical geometry. However, we deliberately remain in the domain of validity of hookean linear elasticity. 

Let us recall the exact energy release rate under zero external load from the Boussinesq problem~\cite{Maugis1995}: 
\begin{equation}
\mathcal{G}=\frac{3K}{8\pi a}\left[\frac{R_0}{2}-\frac{R_0^{\,2}+a^2}{4a}\ln\left(\frac{R_0+a}{R_0-a}\right)\right]^2\ .
\end {equation}
Then, using Eq.~\mref{elasfinbis}, one obtains:\small
\begin{equation}
\label{elasfinter}
U_{\textrm{el}}=\frac{3}{4}KR_0^{\,3}\int_0^{a/R_0}dx\ \left[\frac{1}{2}-\frac{1+x^2}{4x}\ln\left(\frac{1+x}{1-x}\right)\right]^2\\ .
\end {equation}\normalsize
Finally, according to Eq.~\mref{total}, the total energy is thus given by the sum of Eqs.~\mref{workad,tense1,elasfinter}:
\begin{subequations}
\begin{align}
U_{\textrm{TEA}}=-\pi Wa^2+2\pi\gamma \left(a^2+\frac{h^2}{2}\right)\nonumber\\
+\frac{3}{4}KR_0^{\,3}\int_0^{a/R_0}dx\ \left[\frac{1}{2}-\frac{1+x^2}{4x}\ln\left(\frac{1+x}{1-x}\right)\right]^2\ ,
\label{TEAfullbis}
\end{align}
\end{subequations}
where $h(a)$ results from the combination of Eqs.~\mref{radius,contact}. Dividing Eq.~\mref{TEAfullbis} by $2\pi WR_0^{\,2}$, as in Eq.~\mref{utildetea}, allows to get the dimensionless expression of the total energy:
\begin{subequations}
\begin{align}
\tilde{U}_{\textrm{TEA}}= -\frac{\tilde{a}^2}{2}+X\left(\tilde{a}^2+\frac{\tilde{h}^2}{2}\right)\nonumber\\
+\frac{3Z}{2\pi}\int_0^{\tilde{a}}dx\ \left[\frac{1}{2}-\frac{1+x^2}{4x}\ln\left(\frac{1+x}{1-x}\right)\right]^2\ ,
\label{TEAfull}
\end{align}
\end{subequations}
where we have introduced the dimensionless quantities:
\begin{subequations}
\begin{align}
\label{htil}
\tilde{h}&=\frac{h}{R_0}\\
\label{atil}
\tilde{a}&=\frac{a}{R_0}\ ,
\end{align}
\end{subequations}
according to Eqs.~\mref{radius,contact,defo}.

First, in the limit of no elasticity, one sets $Z=0$ in Eq.~\mref{TEAfull} and the equilibrium depends only on $X$. Minimizing Eq.~\mref{TEAfull} with respect to $Y$ leads to the exact solution:
\begin{equation}
\label{yd1}
Y^*_{\textrm{YD}}=1-\left(\frac{X-\frac{1}{2}}{X+1}\right)^{1/3}\ .
\end{equation}  
According to Eq.~\mref{X}, $X$ is given by:
\begin{equation}
\label{spread}
X=\frac{1}{2+S/\gamma}\ ,
\end{equation}  
where we introduced the spreading parameter (see Eq.~\mref{W}):
\begin{subequations}
\begin{align}
S&=W-2\gamma\\
&=\gamma_{\,\textrm{SV}}-\gamma_{\,\textrm{PS}}-\gamma\ .
\end{align}
\end{subequations}  
We can immediately check that our problem of balance between adhesion and surface tension is defined only if $X>1/2$, that is for partial wetting: $S<0$. Otherwise, when $X\rightarrow1/2$ (or $S\rightarrow0$), one has $Y^*\rightarrow Y^*(1/2)=1$, that is total wetting. Let us now introduce the contact angle $\theta$. Using spherical geometry (see Fig.~\ref{fig1}), we have the relationship:
\begin{equation}
\cos\theta=1-\frac{h}{R_1}\ ,
\end{equation}  
which can be rewritten using Eqs.~\mref{radius,htil}, as follows: 
\begin{equation}
\cos\theta=\frac{4-2\tilde{h}^3}{4+\tilde{h}^3}\ .
\end{equation}  
Then, changing variables through Eqs.~\mref{defo,Y} leads to:
\begin{equation}
\label{cos}
\cos\theta=\frac{1-4(1-Y)^3}{1+2(1-Y)^3}\ .
\end{equation}  
One can incorporate the solution given in Eq.~\mref{yd1} in Eq.~\mref{cos} and obtain the solution $\theta^*$ through:
\begin{equation}
\cos\theta^*=\frac{1-X}{X}\ .
\end{equation}  
Finally, according to Eq.~\mref{spread}, we find:
\begin{equation}
\cos\theta^*=1+\frac{S}{\gamma}\ ,
\end{equation}
which is identical to Eq.~\mref{it} and thus to Young-Dupr\'e law~\cite{deGennes2003}.

\begin{figure*}[t!]
\centering
\includegraphics[width=9.cm]{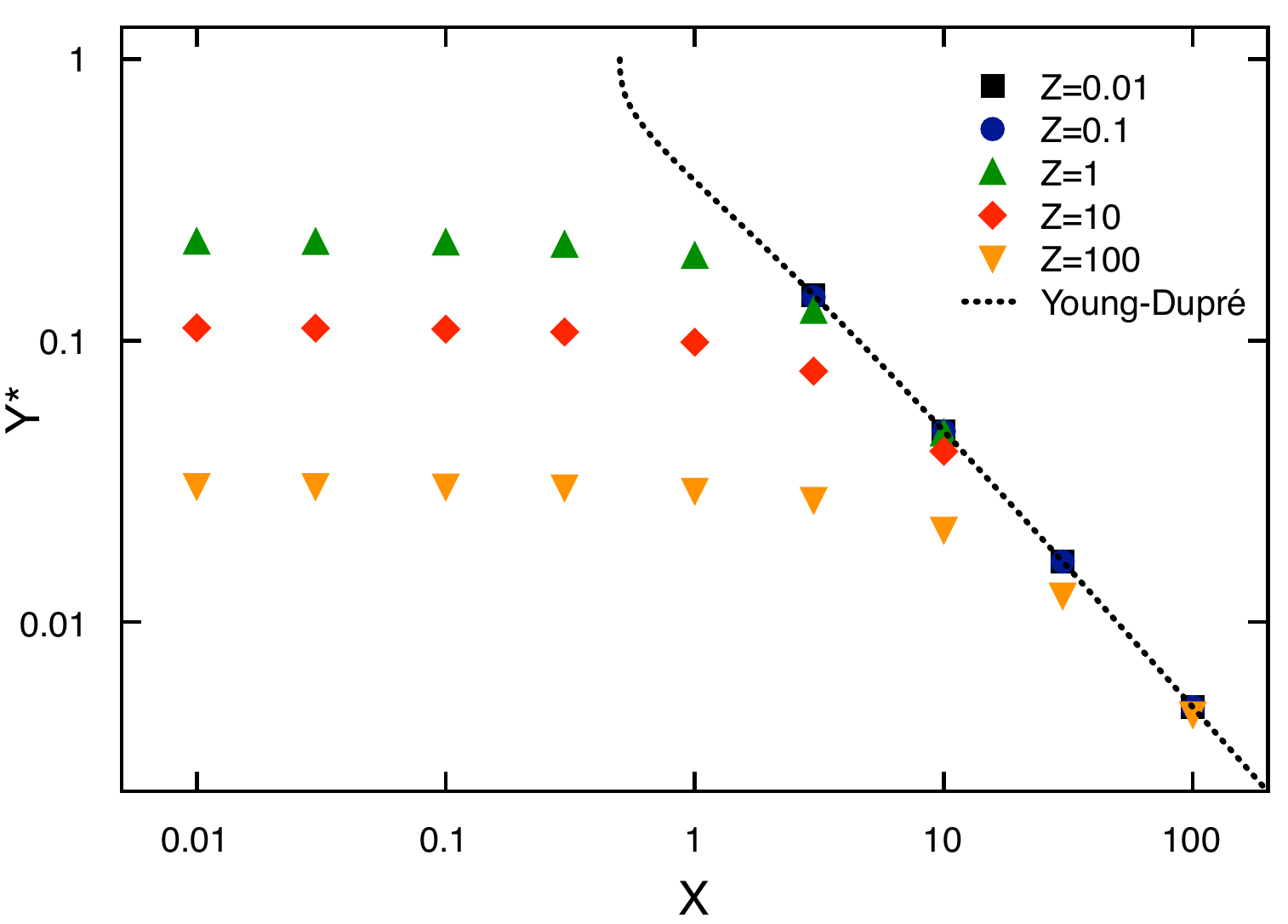}
\includegraphics[width=9.3cm]{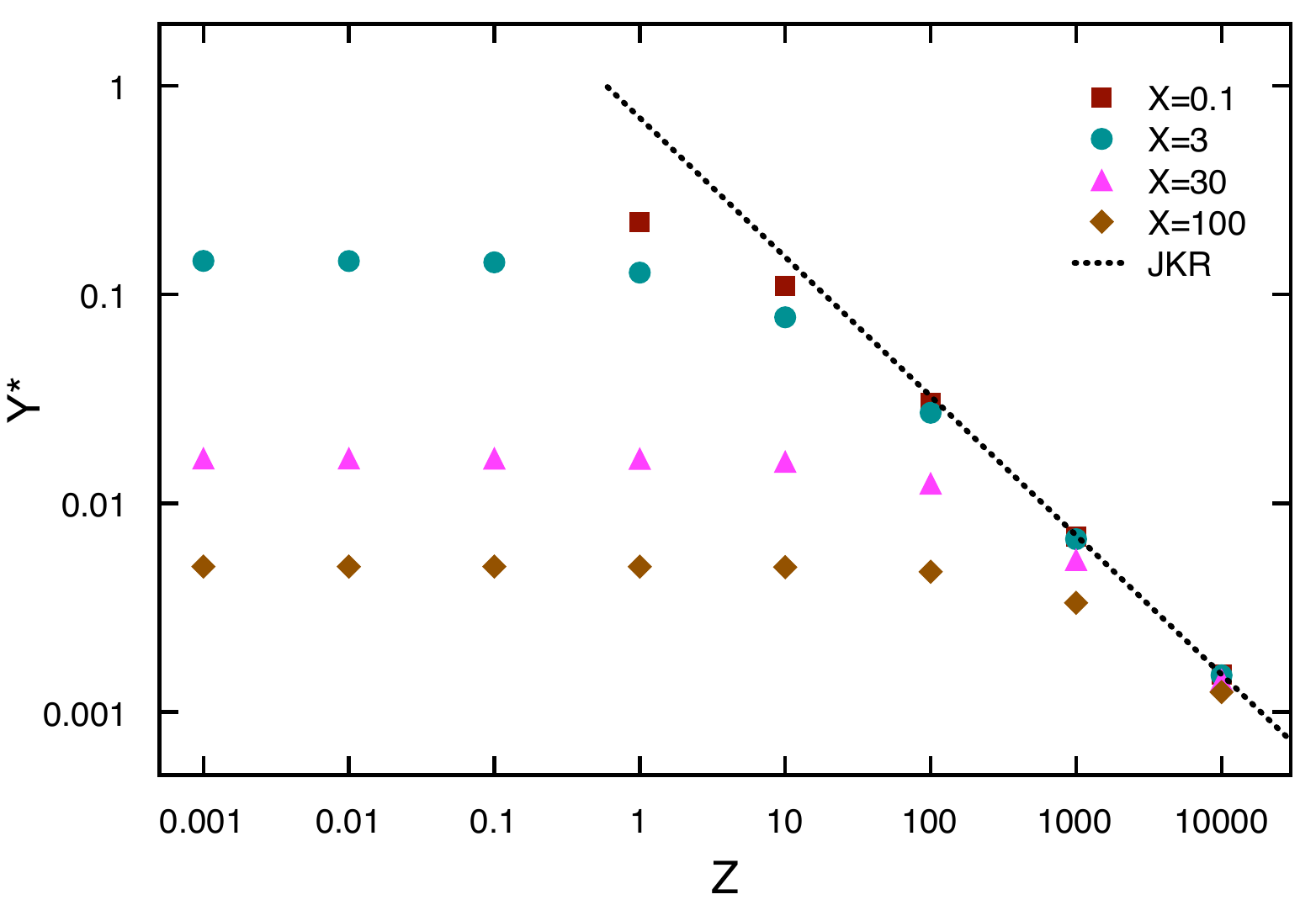}
\caption{\textit{Projections of the numerical solution $Y^*(X,Z)$ of the TEA model at large deformation, from Eq.~\mref{sol}. For comparison, we plotted the Young-Dupr\'e regime ($Z=0$), according to Eq.~\mref{yd1}, and the small deformation JKR regime ($X=0$) according to Eq.~\mref{jkrsmall} with restriction to the $Y^*<1$ domain. Note the singular behaviour of the Young-Dupr\'e model at the transition to total wetting ($X=0.5$). The definitions of the dimensionless variables $X$, $Y$ and $Z$ are given in Eqs.~\mref{X,Y,Z}. }}
\label{fig3}
\end{figure*}

Let us now study the general case. According to Eqs.~\mref{radius,contact,defo,TEAfull,htil,atil}, the dimensionless energy $\tilde{U}_{\textrm{TEA}}(X,Y,Z)$ is now a function of one variable $Y$ that describes the deformation, and two physical parameters: $X$ that quantifies capillarity over adhesion, and $Z$ that quantifies elasticity over adhesion. At given parameters $X$ and $Z$, equilibrium is reached for a minimum of $\tilde{U}_{\textrm{TEA}}$ with respect to $Y$. Thus, the solution $Y^*$ satisfies:
\begin{equation}
\label{sol}
\left(\frac{\partial\tilde{U}_{\textrm{TEA}}}{\partial Y}\right )_{Y=Y^*}=0\ .
\end{equation}  
In contrast to the small deformation problem, where the solution could be expressed as a function of a unique variable, here the solution $Y^*(X,Z)$ is a 2D surface. This double dependence on $X$ and $Z$ at large deformation could be an explanation of the spreading of the numerical simulation data in previous work~\cite{Carrillo2010}. We solved Eq.~\mref{sol} numerically for several couples of parameters $X$ and $Z$. Projections are shown in Fig.~\ref{fig3}.
As expected, at small $Z$ or large $X$ one recovers the Young-Dupr\'e regime, and at small $X$ or large $Z$ (and thus small deformation $Y^*$) one recovers the JKR regime. The results are also in good agreement with numerical simulations~\cite{Carrillo2010} at small adhesion parameter.
\bigskip

\subsection*{Conclusion}  
We reported on a complete model for Tenso-Elastic-Adhesive (TEA) spheres placed on a rigid substrate, for both small and large deformation cases. Interestingly, the small deformation case offers an exact analytical solution that connects the JKR and Young-Dupr\'e asymptotic regimes through a single parameter dependence. We thus predicted a condition to observe this crossover experimentally: $\gamma^{\,3}\approx WK^{2}R_0^{\,2}$. Moreover, using an analogy with fracture mechanics that was originally proposed by Maugis, we clarified the difference with previous models in the literature. The large deformation energy was then obtained through this analogy with fracture mechanics, and its minimization led to equilibrium with a double parametric dependence. This work opens the way to quantitative experiments and numerical simulations on soft particles with large deformation, where the typical elastic $KR_0$, adhesive $W$ and tensile $\gamma$ surface energies are of the same order of magnitude. From an experimental point of view, one may imagine using electrowetting~\cite{Pollack2000,Mugele2005} in order to scan the adhesive parameter independently, and thus probe this striking crossover between adhesion and wetting of soft objects. Non-linear elastic materials may be studied as well through a neo-hookean approach. In near future, this work should also be connected to DMT theory~\cite{Derjaguin1975,Maugis2000}. Finally, viscoelastic dynamical studies may enlarge the scope of the present static analysis to a wide range of experimental situations.

\subsection*{Acknowledgments}
The authors would like to thank \'Ecole Normale Sup\'erieure of Paris and Fondation Langlois for financial support. They also thank Jan-Michael Carrillo and Andrey Dobrynin for sharing numerical data and interesting discussions, as well as Antonin Borot for providing reference 13, and Etienne Rapha\"el for the cover artwork.

\end{document}